\documentclass{article}\textheight 9in \textwidth 6.5in
 \usepackage{graphicx} \usepackage{amssymb}\usepackage{stmaryrd}
 \usepackage{amstext} 

 \newcommand\drp[1]{}

\newcommand\AND{\&}

 \newcommand{\trm}[1]{{\bf\em #1}}
 \newcommand\floor[1]{{\lfloor#1\rfloor}}
 \newcommand\ceil[1]{{\lceil#1\rceil}}
 \newcommand{\tuple}[1]{\langle{#1}\rangle}

 \newcommand\aft{\succ} 

 \newcommand\ie{{i.e.,\ }} \newcommand\eg{{e.g.,\ }}
  \newcommand\inc{{incl.\ }}
 
 \newcommand{\E}{\mbox{\bf E}} \newcommand{\dir}{\mathbf{dir}}
 \newcommand\edf{{\stackrel{\mbox{\tiny def}}=}}

 \newcommand{\qed} {{\hfill\rule{0.8em}{1.8ex}}}
 
 \newcommand\T{\Gamma} \renewcommand\a{\alpha}
  \newcommand{\ov}{\widehat}
 \newcommand\s{\sigma} \newcommand{\x}{\chi} \renewcommand\r{\rho}

 \newtheorem{rem} {Remark} [section] \newtheorem{thm} {Theorem} [section]
 \newtheorem{claim}[thm]{Claim} \newtheorem{cor} [thm] {Corollary}
 \newtheorem{lem} {Lemma} [section] 
 \newcommand {\BT} {\begin{thm}} \newcommand {\ET} {\end{thm}}
 \newcommand {\BL} {\begin{lem}} \newcommand {\EL} {\end{lem}}
 \newcommand {\BCl} {\begin{claim}} \newcommand {\ECl} {\end{claim}}
 \newcommand {\BC} {\begin{cor}} \newcommand {\EC} {\end{cor}}

  \newcounter{action}\newcounter{actionl}

 \newcommand\reborn{{\sf reborn}}
  \newcommand\start{{\sf start}}
  \newcommand\activ{{\sf active}}
  \newcommand\done{{\sf done}}

 \newcommand\adv{{\em Adversary}}

 \newcommand\zhi{{\text{high}}}
 \newcommand\zlow{{\text{low}}}  \newcommand\Zlow{{\text{Low}}}
 \newcommand\sgl{{\text{\footnotesize$\mathord\downarrow$}}}
 \newcommand\dbl{{\text{\footnotesize$\mathord\Downarrow$}}}
 \newcommand\dplx{{\text{\footnotesize$\mathord\downdownarrows$}}}
 \newcommand\sprd{{\text{\footnotesize{$\mathord\downarrow$}\hspace{-1.5pt}\text{\scriptsize$\upharpoonleft$}}}}
 \newcommand\splt{{\text{\footnotesize$\mathord\curlywedgedownarrow$}}}
 \newcommand\stb{{\text{\footnotesize--\hspace{-4pt}$\mathord\downarrow$}}}
 \newcommand\brn{{\text{\boldmath$_\circ$}}}

\newcommand\zon{\sf on}\newcommand\zoff{\sf off}
\newcommand\zlb{\text{\footnotesize$\mathord{\shortuparrow}$}}
 \newcommand\zrst{\text{\zon\zlb}}
 \newcommand\zrtON{\text{\zon${\brn}$}}
 \newcommand\zrtOFF{\text{\zoff${\brn}$}}
 \newcommand\zsprd{hook}
 \newcommand\zstub{stub}

 \newcommand\zold{{\text{\sf \zon\dplx}}}
 \newcommand\znew{{\text{\sf \zoff\dplx}}}
 \newcommand\znewp{{\text{\sf \zoff\footnotesize\dbl}}}
 \newcommand\zoldp{{\text{\sf \zon\footnotesize\dbl}}}
 \newcommand\sglOFF{{\text{\sf \zoff\footnotesize\sgl}}}
 \newcommand\sglON{{\text{\sf \zon\footnotesize\sgl}}}

 \newcommand\verbfont[1]{{\em {\sf #1}}}
 \newcommand\zcrash {{\verbfont{crash}}}
 \newcommand\Zcrash {{\verbfont{Crash}}}
 \newcommand\Float {{\verbfont{Float}}}

 \newcommand\grnd {{\verbfont{ground}}}

 \newcommand\zlock {{lock}} 
 \newcommand\zlocks{{locks}} 
 \newcommand\zopened{{\sf open}} 

 \newcommand\zund{ {\#} } 

 \newcommand\zgrnd {{root$_{\text{\tiny\Si}}$}} 

 \newcommand\zrise{{\em rise}}

 \newcommand\Grd{{\text{$\cal G$}}}
 \newcommand\Si{{\text{\bf R\/}}} 
 \newcommand\Se{{\text{\bf Sh\/}}}
 \newcommand\Fnc{{\text{\bf F\/}}} 
\newcommand\Le{{\text{\bf LE\/}}}
 \newcommand\LEL{{Leader Elector}}
 
\newcommand\bfs{{\text{\sc d\/}}} 
 \newcommand\AM{{\text{\sc c\/}}} 
 \newcommand\CC{{\text{\sc cc\/}}} 
 \newcommand\CB{{\text{\sc cm\/}}} 

 \newcommand\chk{\Lambda}
 \newcommand\slv{\Phi}

 \newcommand\Env{\cal E}

 \newcommand\jj{{\ov{\jmath}}}

 \newcommand{\up}{{\mbox{\bf up}}\newcommand{\UP}{\mbox{\bf Up}}}

 \newsavebox\hth\sbox\hth{{\protect\mbox{\sf h{\scriptsize\rm 3}}}}
 \newcommand\zh{\protect{{\usebox\hth}}}

 \newcommand\sbs[2] {{\text{$\mathsf {#1_{\text{\scriptsize #2}}}$}}}
 
 \newcommand\tle {{\sbs t\Le}}
 \newcommand\tsi {{\sbs t\Si}}
 \newcommand\tFnc {{\sbs t\Fnc}}
 \newcommand\bFnc {{\sbs b\Fnc}} 
 \newcommand\tcc {{\sbs t\CC}}
 \newcommand\tcb {{\sbs t\CB}}
 \newcommand\tbfs {{\sbs t\bfs}}

 \newcommand\dbfs {{\sbs d\bfs}}

\newcommand\hsi{{h_{\Si}}}
\newcommand\hsii[1]{{h_{\Si}^{(#1)}}}

 \newcommand\zp[1]{{p_{#1}}}
 \newcommand\zP[2]{{#2.\zp{#1}}}
 \newcommand\lp[1]{{{\cal L}_{#1}}}

 \newcommand\zpr[1]{{p_{#1}}} 

 \newcommand\zprb{{\zpr {\bfs}}}
 \newcommand\zprB{\vec{\zprb}} 
 \newcommand\zprc{{\zpr {\AM}}}

\newcommand\zbl{b_l}

 \newcommand\bvec[1]{\vec{#1}\!\mathring{}\,}

 \newcommand\Thue{\mu}
 \newcommand\lnd{\lambda}

\def\version {\thanks {A preliminary version of this article appeared
in \cite {STACS}.}}

\begin{document}\title{\null\vspace{-3pc}
 Flat Holonomies on Automata Networks\version~\thanks
 {Supported in part by NSF grants CCR-0311411 and CCR-0311485.}
\\{\normalsize\bf Please take the latest version from}
 {\normalsize\bf http://arXiv.org/abs/cs.DC/0512077 } }
 \author {Gene Itkis\thanks {Boston University, Department of
 Computer Science, 111 Cummington St., Boston, MA 02215.}
 \and Leonid A. Levin\footnotemark [3]
 \date{\em \today }
 } \maketitle

\begin{abstract} We consider asynchronous networks of identical finite
(independent of network's size or topology) automata. Our automata drive
any network from {\em any} initial configuration of states, to a coherent
one in which it can carry efficiently any computations implementable
on synchronous properly initialized networks of the same size.

A useful data structure on such networks is a partial orientation
of its edges. It needs to be flat, \ie have null holonomy
(no excess of up or down edges in any cycle). It also needs
to be centered, \ie have a unique node with no down edges.

There are (interdependent) self-stabilizing asynchronous finite automata
protocols assuring flat centered orientation. Such protocols may vary in
assorted efficiency parameters and it is desirable to have each
replaceable with any alternative, responsible for a simple limited task.
We describe an efficient reduction of any computational task to any such
set of protocols compliant with our interface conditions. \end{abstract}

\section {Introduction} \label {s:intro}

\subsection {Dynamic Asynchronous Networks with Faults}

The computing environment is rapidly evolving into a huge global
network spanning scales from molecular to planetary and set to
penetrate all aspects of life. It is interesting to investigate when
such diverse complex unpredictable networks ---including tiny and
unreliable nodes--- can organize themselves into a coherent computing
environment.

Let us view networks as connected graphs of identical asynchronous
finite automata and try to equip them with a self-organizing protocol.
The automata have no information about the network, and even no room
in their $O(1)$ memories to store, say, its size, time, etc. They run
asynchronously with widely varying speeds. Each sees the states of its
adjacent nodes but cannot know how many (if any) transitions they made
between its own transitions. The networks must be {\em self-stabilizing},
\ie recover a meaningful configuration if faults initialize their
automata in any combination of states whatsoever.\footnote
 {The faults are assumed {\em transient} \ie self-stabilization is
 achieved after faulty transitions seize. Automata constant size and
 uniformity may help comparing neighbors and cutting edges to dissimilar
 ones. Absence of topology restrictions makes cutting-off persistently
 faulty nodes harmless.}

Such conditions and requirements may seem drastic, but stronger
assumptions may be undesirable for the really ubiquitous networks that
we came to expect. For instance, the popular assumption that each node
grows in complexity with the size of the network, keeps some global
information, and yet preserves reliable integrity, may become too
restrictive (and is certainly inelegant).

So, which tasks and how efficiently can be solved by such networks?
The network's distributed nature, unknown topology, asynchrony,
dynamics and faults, etc., complicate this question. The computational
power of any network with total memory $n$ is in the obvious class
Space$(n)$. In fact, this trivial condition is sufficient as well.

\subsection {Orientation and Computing}

We consider protocols based on {\em orientation} for each directed
edge (up, down, or horizontal) implemented by comparing $Z_3$ values
held in nodes. It is a somewhat stretched transplantation to graphs
of widely used geometric structures, {\em connections}, that map
coordinate features between nearby points of smooth manifolds.
Orientation is a simplest analog of such structures, comparing
relative heights of adjacent nodes.

An important aspect of a connection is its {\em holonomy}, \ie the
composition over each circular path (often assumed contractible, though
in graphs this restriction is mute). Connections are called {\em flat}
if this holonomy is null (identity), for each cycle. For our
orientations this means every cycle is {\em balanced}, \ie has equal
numbers of up and down edges.

Here is an example of utility of flat orientations. (Other types of
connections on graphs might be beneficial for other problems, too.) Some
networks deal with asynchrony by keeping in each node a step counter
with equal or adjacent values in adjacent nodes. Nodes advance their
counters only at local minima. For our model, such counters may be
reduced $\bmod\;3$ when no self-stabilization is required. The change of
their values across edges induces orientation, obviously flat. Faulty
configurations, however, can have inconsistent $\bmod\;3$ counters with
{\em vortices}, \ie unbalanced (even unidirectional in extreme cases) cycles.

Flat orientations are especially useful when {\em centered}, \ie having
a unique node with no down edges. It then yields a BFS tree, maintaining
which is known to self-stabilize many network management protocols.

Assuring these properties is the task of our automata. Their constant
size combined with network's permissiveness, present steep challenges,
require powerful symmetry-breaking tools, such as {\em Thue sequences}
\cite{Thue-12} and others. These tools are highly interdependent: each
can be disrupted by adversarial manipulation of others. This makes them
hard to analyze, optimize, and implement.

Here we efficiently reduce these (and thus any other) tasks to several
smaller problems; each can be solved completely independently as long as
the protocols conform to a simple interface preventing them from
disrupting each other. Such protocols may vary in assorted efficiency
parameters, and it is desirable to have each replaceable with any
alternative solving a simple limited task.

\subsection {Maintaining Flat Centered Orientation}\label {s:i-sm}

 The task of assuring a non-centered flat orientation is easier in
some aspects, \eg it can be done deterministically. This is known to
be impossible for the other task, centering an orientation. A fast
randomized algorithm for it, using one byte per node, is given in
\cite {IL92}. The appendix there gives a collection of deterministic
finite automata protocols that make orientation flat, running
simultaneously in concert with each other and with the centering
protocol.

In this paper we refer to three separate tasks:
 (1) rectify orientation on graphs spanned by forest of such trees,
 (2) center such an orientation merging the forest into a tree, and
 (3) fence vortices blocking centering process around them.
 Our main goal is to develop a protocol (4) {\em Shell} that (using no
additional states) coordinates any (\eg provided by an adversary)
protocols performing these four tasks to assure that a centered
orientation is verified and repaired if necessary, with the efficiency
close to that of these supplied underlying task protocols. One more
protocol (5) then efficiently reduces self-stabilization and
synchronization of any computational task to assuring a centered
orientation. The protocol (5) is described in Sec.~\ref{s:prop}. The
tasks (1)--(3) are formally defined in Sec.~\ref{s:std}, and the {\em
Shell} protocol (4) is presented in Sec.~\ref {s:std}.

\subsection {Self-Stabilizing Protocols}

The concept of {\em self-stabilizing} was pioneered by Dijkstra \cite
{D74} and has since been a topic of much research in distributed
computation and other areas (see bibliography by T.~Herman
\cite{Her-bib}). Self-stabilization for typical tasks was widely
believed unattainable unless nodes are not identical or grow in size (at
least logarithmically) with the size of the network. (See, \eg\cite
{MOOY92} for discussion of undesirability of such assumptions.)

Logarithmic lower bounds for self-stabilizing leader election on rings
\cite {IsraeliJ90:rand} (see also \cite {DGS96}) reinforced this belief.
However, such lower bounds depend on (often implicit) restrictions on
accepted types of protocols: configurations with no potential leaders
(tokens) must disappear in one step. Awerbuch, Itkis, and Ostrovsky
\cite {aio-talks}, gave randomized self-stabilizing protocols using
$\lg\lg n$ space per edge for leader election, spanning tree, network
reset, and other tasks. This was improved to constant space per node for
all linear space tasks by Itkis in \cite {aio-talks}, and by \cite
{IL92} (using hierarchical constructions similar to those used in other
contexts in \cite {Thue-12, Ro71, G86}). These results were later
modified in \cite {AO94} to extend the scope of tasks solvable
deterministically in $O(\log^*n)$ space per edge (beyond forest/orientation
construction, for which algorithms of \cite {IL92} were already
deterministic).

There is extensive literature on self-stabilization and similar features
in other contexts which we cannot review here. For instance, many
difficult and elegant results on related issues were obtained for
cellular automata (see, \eg\cite {G86}) on grids. However, the irregular
nature of our networks presents different serious complications.

\section {Models} \label {model}

Our {\em network} is based on a reflexive undirected (\ie all edges have
inverses) connected {\em communication graph} $G{=}(V,E)$ of $n$
nodes, diameter $d$, and degree bound $\Delta$. Nodes $v$ are anonymous
and labeled with {\em states} consisting of bits and pointers to
adjacent nodes $w\in\E(v)$. Protocols are automata operating on
functions of these states called {\em fields}. Their implementation
specifies what changes of states actions on fields imply.

We avoid duplication when an edge carries pointers of several protocols
as follows. The system call creates a {\bf\em hard} pointer and sets a
protocol's {\bf\em soft} pointer to its name. Such soft pointer fields
can be copied by other protocols. Hard pointers are removed when no soft
pointers to them remain. A soft pointer can point at its source node; we
then synonymously refer to it as absent or looping.

A \trm {link} $[v,w]$ is the state of edge $vw$: a network obtained by
renaming nodes $v,w$ canonically and dropping all other nodes;
pointers between $v,w$ (\inc loops) are part of the link. Nodes
\trm {act} as automata changing their states based on the set (without
multiplicity) of all incident links. Thus, a node's state transition may
be conditioned on having (or not) neighbors in some state, but not on
having five of them. When a node sets a hard pointer, it chooses a link,
but not a specific (anonymous) neighbor connected by such a link.
Some protocols may require this choice to be deterministic, \eg using
an ordering of edges. Thus, lemma~\ref{l:rtm} uses it on a tree to
choose each child in turn for the TM simulation.

{\small On a rooted tree with $\Delta=O(1)$, edges can be easily ordered
by parents coloring them in $\Delta$ colors. Then, a general network $N$
with a centered orientation allows a TM simulation by theorem~\ref
{rsp:l}. Such TM can use $\Delta^2$ colors to color distinctly any nodes
with common neighbors, thus ordering each node's edges in $N$. For
non-constant $\Delta$, cyclic ordering of node's edges needs to be
provided by the model.}\footnote
 {For general undirected graphs, cyclic ordering of the edges for each
node is equivalent to embedding the graph in a two-dimensional
orientable manifold.}

\subsection {Asynchrony} \label{s:asynch}

Asynchrony is modeled by {\adv} selecting the next node to act: she
adaptively determines a sequence of nodes with unlimited repetitions;
the nodes act in this order. A network's (or protocol's $P$) \trm {step}
is the shortest time period since the end of the previous step within
which each node acts (or $P$ is called in it) at least once. By
$\tau\aft s$ we denote that all of the step $s$ occurs before the time
instant $\tau$. For simplicity, we assume that only one node acts at any
time. Since node transitions depend only on its set of incident links,
this is equivalent to allowing {\adv} to activate simultaneously any
independent set of nodes.

{\small We could relax this model to \trm {full asynchrony} allowing
{\adv} activate {\em any} set of nodes. This involves replacing each
edge $uv$ with a dummy node $x$ and edges $ux$ and $xv$. This change of
the network affects only our structure fields protocols (assuring
centered orientation: see Sec.~\ref {s:pp}), which tolerate any network.
 Node $x$ is simulated by one of the endpoints,
say $u$, chosen arbitrarily, \eg at random. We call $u$ {\em host} and $x$
{\em satellite}; $v, x$ --- {\em buddies}. When activated by {\adv}, a
node first performs its own action and then acts for all its satellites.
Thus, the dummy nodes never act simultaneously with their hosts.

To avoid simultaneous activation of buddies let each node (real or
dummy) have a black or white color, flipped when the node acts (even if
that action changes nothing else). A dummy node $x$ acts only when its
color is opposite to its buddy's; a real node $v$ acts only when its and
all its buddies' colors match. If a node does not act, in one step its
buddies will have the color freeing it to act. Thus, at the cost of
using a bit per edge, any structure protocol
designed for our model can be run on a fully asynchronous network.}

\subsection {Faults}

The \trm {faults} are modeled by allowing {\adv} to select the initial
state of the whole network. This is a standard way of modeling the
worst-case but transient, ``catastrophic'' faults. The same model
applies to any changes in the network: since even a non-malicious local
changes may cause major global change, we treat them as faults.
After changes or faults are introduced by {\adv}, the network takes some
time to stabilize (see Sec.~\ref {s:pp} for the precise definitions) ---
we assume that {\adv} does not affect the transitions during the
stabilization period, except by controlling the timing (see
Sec.~\ref {s:asynch} above). Our protocols in this paper are all
deterministic and make no assumptions about computational powers of
{\adv}. They may interact with or emulate other algorithms,
deterministic or randomized. These other algorithms may impose their own
restrictions on {\adv}, which would be inherited by our simulations.

\subsection {Orientation and Slope Bits} \label{s:slope}

Edge \trm {orientation} {$\dir()$} of $G$ maps each directed edge $vw$ of
$G$ to $\dir(vw)\in\{0,\pm1\}$.
The \trm {{\zrise}} of a path $v_0\ldots v_k$ is $\sum_{i=0}^{k-1}
\dir(v_{i} v_{i+1})$.
 We consider only orientations for which the {\zrise} of any cycle is
$0\pmod3$. They have economical {\em representations}: Let each
node $v$ keep a \trm {slope bits} field $v.\zh{\in}\{0,\pm 1\}$ and define
$\dir(vw)\edf-\dir(wv)\edf (w.\zh{-}v.\zh\bmod 3){\in}\{0,\pm1\}$.
 We say that $w\in \E(v)$ is \trm {over} $v$ (and $v$ is \trm {under} $w$)
if $\dir(vw){=}+1$; directed edge $vw$ points \trm {up} and $wv$
\trm {down}; define $\up(vw)\edf (\dir(vw){=}+1)$. A path $v_0\ldots v_k$ is
an \trm {up-path} if $v_{i+1}$ is over $v_{i}$ for all $0{\le} i {<}k$.
Cycles of $0$ {\zrise} are called \trm {balanced}, others ---
 \trm {vortices}.

A unique node with no down edges is called the \trm {center}.
We will mark potential centers, calling them \trm {roots}.
We call \trm {flat} an orientation with roots,
each with $\zh=-1$, only up edges, and {\zrise}$\ge0$ outgoing paths.
This implies no vortices and no up-paths\footnote
 {Such paths determine delays in many applications, but higher limits
often suffice. Many algorithms modify orientation gradually, changing
{\zrise} of any path by at most 1 at a time. Then the {\zrise} of any
cycle (being a multiple of 3) stays constant. This limits the cumulative
{\zrise} change of any path to $\pm2d$. Thus, the maximum node-length of
up-paths can vary with time by at most a $2d$ factor.}
 of $>d$ nodes, but is more restrictive than in the Introduction
(Sec.~\ref{s:intro}). A flat orientation with a center is called \trm
{centered}.

\subsection {Tree-CA Time and TM Reversals}

We characterize in usual complexity terms the computational power of
asynchronous dynamic networks $G$ in two steps. First we express it in
terms of Cellular Automata $H$ on $G$-spanning trees ({\em tree-CA}).
 We treat $H$ as a special case of our networks when they are trees
initialized in a blank state and acting synchronously. $H$ holds the
network topology as adjacency lists $l_v$ (say, by the dfs numbering of
the tree) of its nodes $v$. $l_v$ are held in read-only \trm {input
registers}; $v$ have access to one bit of $l_v$, rotated synchronously
by the root.

Once its flat orientation stabilizes, our network can simulate tree-CA
(subsection~\ref {s:rsp-pf}). Tree-CA are simpler than our networks, but
still have significant variability depending on the topology of the
trees. To avoid this variability, we further compare them in
computational power to Turing Machines (TM). Tree-CA can simulate TMs
and vice versa (subsection~\ref {s:CA-TM}). The efficiency of this
mutual simulation seems best expressed using the number of \trm
{reversals} \ie changes of the TM head direction as (parallel) time
complexity. When using this measure \cite {Tr-TMrev,Ba-TMrev}, we refer
to TM as {\em reversal TM (rTM)}.

Our rTM has read-write work and output tapes $W,O$ of size
$\|W\|=\|O\|=n$, and a read-only input tape $I$. For simplicity we
assume rTM's heads turn only when the work head is at the end of its
tape. The bits of tree-CA input registers are stored on rTM's input tape
at intervals $2n$, so that when the work-tape head is in cell $i$, the
input-tape head reads a bit of the $i$'s register.

Ignoring $d,\Delta$ time factors, tree-CA on any tree have the same computing
power as rTM with the same space and time, thus exceeding power of sequential RAM.
rTM can simulate RAM fast but can also, say, flip all bits in one sweep,
which takes $\theta(n)$ RAM time. Variant connectivity gives some
networks greater power of parallelism than others. For instance, tree-CA
take nearly linear time to simulate sorting networks, while the latter
given read-only access to the adjacency list of any other network, can
simulate it (or PRAM) with polylog overhead.

\section {Solving Any Task with Centered Orientation} \label{s:prop}

Consider an rTM algorithm $T_n(x)$ that computes a function $t_n(x)$
when initialized on a working tape of size $n$ with $x$ on the input
tape. $T,t$ are called \trm {constructible} if $T$ runs in (reversal)
time $O(t)$ and space $O(n)$. The running time of any algorithm $T$ is
constructible since $T$ can be modified to count and output its time.

We need to tighten this condition slightly to assure the time bound
 even when $T$ is initialized in maliciously chosen configurations. We
call algorithm $T$, and the function $t_n(x)>\lg n$ it computes, \trm {strictly
constructible} if for some $c\in(0,1)$, $T$ runs in space $O(n/|\lg_c
n|)$ with $O(t^c)$ expected reversals. Most functions $t$ used as time
bounds take for their computation significantly (usually exponentially)
less time and space than $t_n(x)$ steps and $n$ cells. Thus, the
overheads of strict constructibility are rarely an issue.

Let $q$ be an input-output relation on pairs $\tuple{x,y}$ of questions
$x$ and ``correct answers'' $y\in q_x$. With a strictly constructible
time bound $t_n(x)$ it forms a \trm {task} $\T$ if there exist a pair
$\tuple {\chk,\slv}$ of probabilistic algorithms: Checker (needed only if
$\|q_x\|>1$) and Solver, running in space $\|y\|$ and expected time
$t_n(x)$ such that
\begin {itemize}
 \item $\chk_n(x,y)$ never rejects any $y\in q_x$, but with probability
 $>1/2$ rejects every $y\not\in q_x$;
 \item $\slv_n(x)$ with probability $>1/2$ computes $y\in q_x$.
\end {itemize}

Our goal is for any task (specified for a faultless and synchronous
computational model such as rTM) to produce a protocol running the task
in the tough distributed environment where {\adv} controls the timing
and the initial state of the system. We separate this job into two: First,
we assume that some special \trm {structure protocols} generate a
centered orientation and stabilize, \ie the orientation stops changing.
Section~\ref {s:prop} and its Theorem~\ref{rsp:l} discuss how to achieve
our goal after that. The remainder of the paper starting with
Sec.~\ref{s:std} describes the structure protocols, which run in the
special {\em structure} fields.

\subsection {Self-Stabilization} \label{s:pp}

Let each processor (node) in the network $G$ have read-only \trm {input}
field, and read/write \trm {work, output}, and \trm {structure} fields.
 A \trm {configuration} at time instant $\tau$ is a quintuple $\langle
G,I,O_\tau,W_\tau,S_\tau\rangle$, where functions $I,O_\tau, W_\tau,
S_\tau$ on $V$ represent the input, output, work and structure fields
respectively. The structure protocols serve to maintain the centered
orientation. They run in $S_\tau$, are independent of the task and
computation running in $W_\tau,O_\tau$, and affect it only via setting
the orientation fields of $S_\tau$ which the computation can read.

Let $q$ be a set of correct i/o configurations $\langle (G,\!I),\!O
\rangle$, and $\T=\langle T,q\rangle$ be a corresponding task. A
protocol \trm {solves $\T$ with self-stabilization in $s$ steps} if
starting from any initial configuration, for any time $\tau\aft s$ the
configuration $\langle (G,I),O_\tau\rangle\in q$. For randomized
protocols we measure the expected stabilization time. Our protocols do
not halt, but after stabilization their output is independent of the
subsequent coin-flips. (For synchronized protocols stabilization could
also include repetition of the configuration.)

 Protocols, which accept (potentially incorrect $\langle (G,\!I),O'
\rangle\not\in q$ ) halting configurations, cannot be self-stabilizing:
the network put by {\adv} in an incorrect halted configuration cannot
correct itself. Our protocols for $\T$ repeatedly emulate checker
$\chk$, invoking $\slv$ when $\chk$ rejects an incorrect configuration.
We use here the Las Vegas property of (properly initialized) $\chk$: it
never rejects a good configuration. {\adv} may still start the network
in a bad configuration from which neither $\slv$ nor $\chk$ recover
within the desired time. To handle this, we use the self-stabilizing
timer $T$ constructed in Lemma~\ref {l:robust}.

\begin{rem}[Dynamic Properties]\label{rem:temp} For simplicity, we focus
on ``static'' problems. However, the dynamic behavior of protocols is
often of interest as well. We note that many temporal properties can be
achieved by creating (with self-stabilization) a static configuration
that, once correctly established, allows regular algorithms (without
self-stabilization or asynchrony resistance) to assure the desired
behavior. \end{rem}

\BT\label{rsp:l} Any task $\T$ can be solved on any asynchronous
networks $G$ with (unchanging) centered orientation in their $S$-fields by
protocols self-stabilizing in $T(G,I)O(d\Delta\lg n)$ steps.\ET

For a proof we define a \trm {stably constructible} rTM $T_n(x)$ (or
\trm {timer}) as one that starting from {\em any} configuration on
$n$-cell work tape, stabilizes with $O(T_n(x))$ expected time.

\BL\label{l:robust} Any strictly constructible function $t$ can be
computed by a stably constructible algorithm.\EL

When $T_n(x)$ is a timer, any task can be self-stabilized. $M$ keeps two
counters $t,r$ and runs $T$ repeatedly. Whenever $T$ halts, its output
overwrites $t$. Each step, $r$ is decremented if $r\in[1,t]$. Otherwise,
$r$ is reset to $t$ and $M$ runs $\chk$, properly initialized. If $\chk$
rejects, $M$ runs $\slv$. If outputs of $\slv$ are unique, no $\chk$ is needed:
$\slv$ is run always but its rewriting correct outputs makes no changes and
does not disrupt the stabilization.

\paragraph {Proof of Lemma~\ref {l:robust}} Let $C=\ceil{1/(1-c)}$; we
round $c$ to $1-1/C$. First, we set a $\ceil{\lg n}$ steps rTM timer. It
sweeps the tape, each time marking every second unmarked cell. When all
are marked, it unmarks the tape, and restarts. With it, we stabilize the
following $O(k)$ steps task. It computes $k=\ceil{\lg n-\lg(C\lg n)}
$ similarly to the above timer, and by $k$ merges divides the tape into
numbered segments $s_i$ of length $2^k$ ($s_0$ may be shorter), each
keeping a binary counter $r_i$ bounded by $t_i$ with $\|t_1\|=C$,
$\|t_{i+1}\|=\floor {\|t_i\|/c}\approx c^{-i}$).

In each $s_i$, rTM runs $T(x)$ (iterated to error probability $<1/3k$ if
randomized), in parallel. The $i$-th run goes for $t_i$ steps and
restarts from the blank state. If it halts, all other runs are
restarted, too. Thus, if $T(x)$ takes $T_x\in(t_{i-1},t_i]$ steps, then
starting from any configuration, within $t_i<T_x^{1/c}<T(x)$ steps the
$i$-th run restarts from blank state and halts in $<T(x)$ expected time.
\qed

\subsection {Tree-CA, rTM, and Network Simulations} \label{s:CA-TM}

In this section, we consider how tree-CA $H$ and an rTM $M$ can simulate
each other. Let $H$ have $n$ nodes and $M$ have $2n$ cells, numbered
from left to right. We map each node $x$ of $H$ to two cells of $M$,
denoted $x_($ and $x_)$ reflecting the two visit times of dfs traversal
of $H$. Let input tape bits $M$ reads when its work head is at nodes
$x_(,x_)$ and bits in the input register of $x$ reflect each other.
 Let functions $h,g_),g_($ map the tape characters of $M$ to the
automaton states of $H$ and vice versa. We say a machine $A$ {\em
simulates $B$ with overhead $t$} if after any number $i$ of steps (or
sweeps) of $B$ and $ti$ steps of $A$, the state of each cell (or node)
of $B$ is determined by the function $h$ or $g$ applied to the
corresponding node of $A$.

\BL \label{l:rtm} Any tree-CA $H$ (diameter $d$, degree $\Delta$) and
rTM $M$ with matching inputs, can simulate each other: $H$ with
overhead $O(d\Delta)$ and $M$ with $O(d)$.\EL

\paragraph {Proof:} {\bf $H$ simulating $M$.} The automata nodes $x$ of
each depth in turn, starting from the leaves, compute the transition
function $f_x$. This $f_x$ depends on the current states and inputs of
the subtree $t_x$ of $x$ and its descendants. It maps each state in
which $M$ may enter $t_x$ from the parent of $x$ (sweeping the tape
along the dfs pass of $H$) to the state in which it would exit back to
the parent. Once $f_y$ is computed for each child $y$ of $x$, the new
states of $x_),x_($ and $f_x$ are computed in $O(\Delta)$ more steps.
Since the depth of the tree is $d$, it takes $O(d\Delta)$ to compute
$f_{root}$, and thus to simulate one sweep of $M$ work tape.

{\bf $M$ simulating $H$.} Each node $x$ of $H$ corresponds to a pair
$x_($, $x_)$ of matching parentheses enclosing images of all its
descendants (in $t_x$). On each sweep $M$ passes the information between
matching parentheses of certain depth. Nodes $x$ at this depth are
marked as {\em serve}, their descendants as {\em done}, and their
ancestors as {\em wait}. When the root is {\em done}, all marks are
turned to {\em wait} and $M$ starts simulating the next step of $H$
(from the leaves). When $x_($ and $x_)$ {\em wait} and their children
{\em serve}, $M$ serves $x_(,x_)$ as follows.

The next sweep carries the state of $x$ to its children allowing them to
finish their current transition and enter {\em done}. The same sweep
gathers information from the children of $x$ for the transition of $x$
and carries it to $x_)$. The return sweep brings this information to
{$x_($}; at this point, $x_(,x_)$ go into {\em serve} state --- only the
parent of $x$ information is needed to complete the transaction of $x$.

$M$ keeps two counters: for the input register place all automata of $H$
read at this simulation cycle, and for the segment of input tape $M$
reads at this sweep. $M$ reads its input when the counters match. \qed

\paragraph {Proof of {Theorem~\ref{rsp:l}}\label {s:rsp-pf}}
 A centered orientation on $G$ yields a spanning bfs tree via its up
edges. Consider a tree-CA $H$ on it. It can be synchronized by
keeping a second orientation, incrementing its slope bits and making a
step in each node with no tree-neighbors under it. $H$ in turn
emulates an rTM $M$. We also need $G$ to simulate the rotating
registers of $H$ carrying addresses of their $G$-neighbors.

The vertices are numbered linearly on the tape of $M$ covered with
counters, each with the number of its first vertex. Such counters are
initialized in $O(\lg n)$ time similarly to marking the intervals in
Lemma~\ref {l:robust} proof. The root keeps a (rotating) place $i$ and
all points display the $i$-th digit of their numbers, giving access to
it to all network neighbors. An adjacency list look-up can thus be
simulated in $O(d\Delta\lg n)$.\qed

\section {Assuring Centered Orientation: Problem Decomposition}
\label {s:std}

The protocols in Theorem~\ref{rsp:l}, use centered orientation (in {\zh}
fields, Sec.~\ref{s:slope}). The rest of the paper reduces assuring such
an orientation to three separate tasks of: orientation Rectifier $\Si$,
{\LEL} $\Le$, and Fence $\Fnc$ blocking $\Le$ around vortices. This
section presents these tasks in terms of interfaces (read/write
permissions for fields a protocol $P$ shares with its \trm {environment}
$\Env_P$) and commitments (with time parameters $\tsi,\tle,
\tFnc$). Any protocols complying with these {\em contracts} will work for our
reduction,
 given below as the {\bf Shell} protocol $\Se$. 
$\Se$ uses only one bit $\bFnc$ and one pointer $\zp b$ (it also reads
 pointer $\zp l$).\footnote   
 {The tasks of $\Si$ and $\Fnc$ correspond roughly to the two functions
of {\em SI} in \cite{IL92} -- initiating a flat slope and keeping nodes
open for $\Le$. While \cite{IL92} protocols comply with our contracts,
they had other interdependences and were not designed to take full
advantage of the efficiencies allowed by the separation provided here by
$\Se$ and contracts. {\em SI} was concerned only with $n^{O(1)}$
time-bounds, while here our $\Se$ preserves the efficiency up to factors
$d^{O(1)}$, possibly exponentially smaller than the number of nodes $n$.
Our present $\Se$, $\Fnc$, and (sketched in the appendix) $\Si$ adjust
{\em SI} tasks to the new opportunities.}

\paragraph {Legality, Guard, and {\Zcrash}ing.} {\adv} initiates the
network with arbitrary links, possibly ``abnormal,'' disruptive for
$P$. Correcting them might be hard for $P$: it is 
restricted by the interface and acts at one node at a time, affecting
all incident links, not just abnormal ones. Let $P$ come with a list of
\trm {$P$-legal} links; $v$ is \trm {$P$-legal} if all links exiting
it are or if $v$ to $\zrst$ (defined below).
Any activated $v$ invokes a function {\em guard} $\Grd$, with the list
of illegal links and access to all fields.
It \trm{crashes} illegal $v$ into {\zrst}, and does nothing else. 
$P$-legality of nodes and in-links must be preserved by {\zcrash} and
any actions $P$ makes or permits to $\Env_P$.

\paragraph {Shell fields.} 
$\Grd,\Si$ (and only they) create roots -- potential centers of the
orientation. $\Le$ ``uproots'' 
them and, in non-roots, calls {\Float} which, with no edges to roots
or down, increments $\zh$.
 Eventually the orientation has a center led to by all down paths. 
Uprooting creates non-root local minima, and thus,
down-paths not leading to roots. To guide to roots, $\Le$ keeps \trm
{lead} pointers $\zP lv{=}\vec{v}$; $\zp l$ \trm{loops}
($\vec{r}{=}r$) in roots, cutting off pointer chains. 
Invoking $\Le$ at $v$, $\Se$ copies $\zP lv$ to the \trm {backup} $\zP
bv$ (to help other protocols adjust  if $\Le$ changes $\zp l$).
 $\Se$ initiates $\Fnc$ on a $\zp l$-tree by turning {\zon} its root's
\trm {fence bit} or \trm {phase} ($r.\bFnc{\gets}1$); $\Fnc$ exits
turning it {\zoff} ($r.\bFnc{\gets}0$; only $\Fnc$ can turn the roots
{\zoff}).\footnote  
 {The fence bit $\bFnc$ is used to pass control between $\Fnc$
 and $\Se$ analogously to the control bit in \cite{IL92}.}

\paragraph{Notation.}
   $\lp l$ (\trm{\zstub} {\stb}), $\lp b$:  $\zp{}$-loop predicates;
  $\zP{bl}v$: $(\zP bv).\zp l$; $\lp{bl}$: $\zP{bl}v{=}v$, etc.
Adjacent {\zstub}s are \trm{\zlock}s, isolated -- \trm{roots}.
  $\lp i$: $i{=}\lp b{+}\lp l{\in}\{0,1,2\}$.
\brn: $\lp 2$;
\trm{single} {\sgl}: $\lp 1 \AND\lp b$; 
\trm{reset} {\zlb}: $\lp 1 \AND\lp l$.
\trm{Duplex} {\dplx} ($\lp 0$) are
 \trm{double} {\dbl} if $\zP bv{=}\vec{v}$, else
 \trm{\zsprd} {\sprd} if {\zoff}$\,\AND\,\lp{bl}\,\AND\,\zP bv{\notin}\dbl$,
 \trm{split} $\splt$ otherwise.
\trm{Ground}: root or $\lp{bl}$ split.
We denote $\Se$ states by $\bFnc$ and pointer pattern
(\eg $\zrtON,\sglOFF$).

\paragraph {Height.\label{s:ht}}
 \trm{Senior} pointer $\zP Bv$ loops in ground, is $\zP bv$ in other
 splits, 
 $\vec{v}$ otherwise. 
The \trm{height} $h(v)$ of $v$ becomes undefined ($\bot$) when $v$
is {\zlock} or {\zcrash}es,
 and remains so until non-{\zlock} $v$ changes $\Se$ field(s).
Otherwise $h(v){\edf} v.\zh$ in ground $v$. 
For other $v{\neq}\zP Bv{=}w$,
 $h(v)$ is $h(w){+}\dir(wv)$, retaining its previous value if
$h(w){=}\bot$. 
 A directed edge $vw$ becomes \trm {bound} when $w$
 or its $\zp l$-descendant changes $\bFnc{\gets}1$,
 or the senior ancestor root of $v$ or $w$ changes between $\zrtON$
 and $\zrst$.
 It reverts to \trm {unbound} when $v$ {\zcrash}es. 
 Around vortices {\zrise} varies with paths and edge ends may differ by
 ${>}1$ in height;  such edges are called \trm {rips}.

\paragraph {Symmetry breaking.} $\Si$ (with minimal help from $\Fnc$) 
maintains a hierarchic structure on trees to enable initiation of
parallel $\Si$ protocols.  It is kept via sign bits $\lnd(k)$ of
$v.\zh{=}\pm0$, where $k{=}h(v)/3 {=}2^i(4j{+}s)$, $s{=}{ \pm}1$ and
$\lnd(k){=}$sgn$(s)$.\footnote 
{ This sequence $\lnd$ is based on one used (implicitly) in \cite
{Ro71}, and discussed in \cite{Le05}. \cite {IL92} uses instead
$\Thue(k)$ (based on \cite{Thue-12}) defined as ``$-$'' if binary
encoding of $k$ has an odd $>1$ number of $1$s, or ``$-$'' otherwise.
}
As an exception, we set $\lnd(k{+}1)$ to $-$, marking ``round'' $k{=}2^i
(4^c{+}j^2),j{<}2^c$ with an otherwise impossible \trm {mark}
pattern ${-}{+}{-}{+}$. Here $c$ is a constant that depends on the one in the
commitment ($\Le$.ht) below.
Any segment of $\lnd$ with two marked heights determines them uniquely.
Thus, $\Si$ can use the slope bits $\zh$ to quickly detect rips even
when the senior chain is much larger than the height. 

\subsection{Protocols}\label{s:prot}
\paragraph{Interface permissions.}
Read restrictions serve only to help reader's focus; write
restrictions apply only to the \trm{shared fields} ($\zh,\zp l,\zp b, \bFnc$).
$\Env$ of each protocol can do all actions of $\Se$,
and (when $\Se$ calls other protocols $P$) those listed below as
permitted to $P$. 
$v$ is \trm {ready} if $\vec v{=}v$ or $v.\bFnc{\ne}\vec v.\bFnc$, or
$v{\in}\dbl, \vec{v}{\in}\zlb$.
$\Si,\Grd$ can {\zcrash} any $v$. Otherwise, shared fields change
only in ready $v$ with no ready $\zp l$-child, and
$\Si$ can change only {\zlocks}
(not to {\zoff} with $\zp l$-children).
$\Fnc$ changes only $\bFnc,\zp b$ in roots and $\zh{=}\pm0$ signs. 
$\Si$ can \trm {open} {\zlock} $v$ into $\sglON$ with
 $\zP bv{\in}\zrst$ under $v$,
 all down and no up edges of $v$ going to {\zstub}s.
 $\Si$ can decrement $\zh$ of {\zlocks} with no up edges to
 non-{\zstub}, and change $\pm0$ sign. 
Only $\Si$ can set {\zoff\zlb}.
$\Le$ reads $\dir(),\zp l$, calls {\Float} and moves
 $\zP lv$ (to ${\ne}v$);
it idles in $v,w$ if $\vec v{=}v$ and $\dir(vw){\ne}1$.

\paragraph {Shell.} 
$\Se$ starts by changing $\zrtOFF$ to $\zrtON$, and
 invoking $\Fnc,\Si,\Grd$; {\zlock}s with $\dbl$-children change to $\zrst$.
Invoked in other (ready) non-locks, $\Se$ does the following.

{\bf Split:} $v$ invokes $\Le,\Fnc,\Si$ if $v$ 
(1) is  {\zrtOFF}, or {\sglON} with $\lp{b}(\vec{v})$,
 has (2) a $\dbl$ or no child,
 (3) no split $\zp b$-child, and (4) no $\sgl$child.
 Before this, $\Se$ sets $\zp b$ to $\zp l$ or, in root, to a \dbl
 child, if any. 
Uprooted childless $v$ turns $\dbl$.

{\bf Merge:} 
Activated as $\splt$, $v$ merges 
(1) into {\dbl} if $\vec{v}{\in}\sglOFF$ and $v$ has $\sgl$ or no
child,
(2) into {\sgl} if $\vec{v}$ is $\dplx$ and
   (a) $v{\in}\zon$ has a $\sgl$ or no child, 
or (b) $v$ has $\zoff$ children, all  $\splt$ or $\sprd$.
 $\dbl$ merges into $\sgl$ if $\vec{v}$ is $\zlb$ or $\zon\dplx$.

{\bf Phase Wave:} 
Then $\Se$ sets $v.\bFnc{\gets}\vec v.\bFnc$,
changing $\sprd$ to $\sgl$, and $\sgl$ with a child and a $\dplx$ parent,
to $\sprd$.

\newcommand\cmt[1]{{\par\noindent\makebox[4pc][l]{ \bf (#1):}}}

\paragraph{Commitments.} \label{s:L-contr}
After the first step (when $\Grd$'s
{\zcrash}es stop) under the above Interface and Shell:\label{s:cnrt}
 \cmt {$\Le$.ht} $\Le$ assures a segment of {\zrise} $c\cdot m$,
               $c{=}\theta(1)$ in any $m$-node $\zp l$-chain.
 \cmt {$\Fnc$.cln} $\Fnc$ assures that no $v$ with $v.\bFnc{=}0
      {\ne}\vec v.\bFnc$ has a $\bFnc{=}0$ $\zp l$-ancestor.
 \cmt {$\Fnc$.sgn} $\Fnc$ sets the sign of $\zh$ to
 $\lnd(h(v)/3)$ in (ready) $v$ with a bound in-edge and $\zh{=}\pm0$.
 \cmt {$\Fnc$.rip} $\Fnc$ assures that senior chains from bound rips
do not change.
 \cmt {$\Si$.stb} With the above commitments, $\Si$ \trm
      {stabilizes} in $\tsi$ steps: {\zcrash}es stop, orientation is flat.
 \cmt {$\Fnc$.off} $\Fnc$ turns each root {\zoff} every $\tFnc$
                   steps after $\Si$ stabilization.
 \cmt {$\Le$.ct} $\Le$ centers orientation within expected $\tle$
               $\Le$-steps after $\Si$ stabilization.

\subsection {Shell Performance}\label{s:sh-pf}

\label{s:upr}
 A non-{\zlock} is \trm{{\zlow}} if it has only {\sgl} and
{\stb} ancestors (\inc self), \trm{{\zhi}} otherwise; a {\zhi} with a
{\zlow} parent is \trm{border}. Only $\sglON$ occurs in
both {\zhi} and {\zlow} (but not border). 
 A node becomes {\zhi} (border) only as a result of invoking $\Le$ in
leaves of {\zlow}.
 A root, after invoking $\Le$ (unless uprooted) resets its tree to
{\zlow} by passing through $\zrst$, and a new cycle of $\Le$ calls starts.
Intuitively, $\Fnc$ waits for the whole $\zp l$-tree to turn {\zon},
checks it for rips (more precisely, $vu$ such that the root-root path
against $\zp l$ pointers, across $vu$, and then along senior chain,
has non-0 variance), and, if none, turns the tree root {\zoff} (then $\Se$
propagates {\zoff} through the tree). 
Turning {\zoff}, double children of a split become single, so the
split merges at the next {\zoff} ---after completing a full $\Fnc$
cycle with its checks. 
However, a split $v$ merges \trm{prematurely} if it has no
children (when turning {\zoff}) or if it has only split children and
$\vec{v}{\in}\zold$ (thus, \eg as a $\zp l$-chain of splits turns {\zon}, the
alternating ones merge prematurely; the remaining splits will merge
upon the next {\zoff} wave).
Uprooting, $r$, if childless, instantly merges into its new tree; if
 with a double child $w$, remains ground (but now a split).

We show that centered orientation will be assured by any protocols
that satisfy the above commitments. 
 For the rest of the subsection assume that $\Si$ has stabilized
($\Si$.stb): the orientation is flat (\inc has roots, no {\zlock}s),
$\Si$ no longer changes any shared fields (and thus can be ignored).
Then any $\zp l$-chain is at most $O(d)$: the orientation flatness
bounds {\zrise} by $O(d)$, and ($\Le$.ht) extends this bound to the
length of $\zp l$-chains. 
For every root $r$, $\Fnc$ changes {\zrtON} to {\zrtOFF} within $\tFnc$ steps
($\Fnc$.{{\zoff}}), and then (unless $r$ uproots) $\Se$ changes it
back to {\zrtON} in one more step (after all its $\zp l$-children had a
chance to copy $r.\bFnc$). Assume $\tFnc{=}\Omega(d)$ (otherwise we may need to
replace $\tFnc$ with $\tFnc+d$ below).
A node $v$ is a \trm {switch} if $v.\bFnc{>}(\vec v).\bFnc$.

{\em For any $v$, $v.\bFnc{=}1$ within $O(d)$ steps.}
Indeed, let $v.\bFnc{=}0$. ($\Fnc$.{cln}) assures any {\zon} $\zp
l$-child of $v$ has no {\zoff} child, in a step all children of $v$
are {\zoff}. The maximal {\zoff} $\zp l$-chain from $v$ gets shorter
within each step.

{\em For any $v$, $v.\bFnc{=}0$ within $O(d\tFnc)$ steps.}
Indeed, a {\zlow} {\zon} $v$ changes to {\zoff} or {\zhi} within $O(\tFnc)$ steps: its
root is turned {\zoff} or uproots (making $v$ {\zhi}) within $\tFnc$ 
($\Fnc$.{\zoff}); if its root is {\zoff}, the {\zon} $\zp l$-chain from {\zlow} $v$
shrinks ($O(d)$ times) within a step till $v$ either splits or changes
to {\zoff}.
After the initial $O(d)$ steps a {\zhi} node does not invoke $\Le$
(an {\sglON} with an $\znew$ parent changes to $\zoff$).
Then for a {\zhi} $v$ consider the maximal {\zhi} {\zon} $\zp l$-chain to a
split. This chain can only shrink if the split changes to {\zoff} (and
then within $O(d)$ so does $v$). Within $O(\tFnc)$ steps the chain
either grows (at most $O(d)$ times) or $v$ changes phase: its nearest
{\zlow} ancestor becomes {\zoff} or {\zhi} within $O(\tFnc)$, either becoming
a split (increasing the chain), or $\znewp$ (and then the {\zon} $\zp
L$-chain from $v$ shrinks each step). Thus, $v.\bFnc{=}0$ within
$O(d\tFnc)$ steps.

{\em A split $v$ can change $v.\bFnc$ $\leq3$ times without
  merging, thus $v$ merges in $O(d\tFnc)$. } 
Indeed, when $v$ changes to $\zon\dplx$ it looses its double
  children (or merges). Then it merges by the next change to {\zoff}. 

{\em  A {\zlow} $v$ invokes $\Le$ within $O(d^2\tFnc)$.}
Indeed, any {\zlow} leaf looses its split $\zp b$-children in
$O(d\tFnc)$
(similarly, if it is a root its existing split children merge into $\dbl$),
 and then invokes $\Le$ the next time it is a
switch (or {\zrtON} with only $\zoldp$ children). The depth of the {\zlow} node
(sub)tree (of $v$) can be so reduced $O(d)$ times.

\BL\label{l:L-inv}
Any node invokes $\Le$ within $O(d^3\tFnc)$ steps.
\EL

Indeed, consider a {\zhi} $v$ and the shortest $\zp l$-chain 
from $v$ to a border, split or ground $w$ (possibly ${=}v$).
 Such a chain cannot shrink without $v$ invoking $\Le$: new grounds
are not created any more (except when a childless root floats possibly
making its new parent a ground) and new splits are created with only
border children. 
Moreover, within $O(d^2\tFnc)$ steps the chain grows or $v$ invokes
$\Le$: If $w$ is a root, then $v$ is {\zlow} (and invokes $\Le$
within  $O(d^2\tFnc)$); otherwise, if $w$ is a split, it merges
in $O(d\tFnc)$; and if $w$ is a double, then in  $O(d^2\tFnc)$ $\vec w$
invokes $\Le$ and $w$ changes to single $O(d\tFnc)$ steps later.
 Since this chain can grow only $O(d)$ times $v$ will invoke $\Le$
within $O(d^3\tFnc)$ steps. $\qed$

Since $\Le$ interface fields are not affected by any other protocols, 
this lemma implies \trm{prompt} (polynomial in the network diameter $d$ and
degree $\Delta$) centralization: 

\BT[Main]
Given any contract abiding protocols $\Le, \Si, \Fnc$, our Shell
$\Se$ assures centered orientation within expected
$\tsi(\Delta,d)+O(d^3\tFnc\tle)$ steps.
\ET

\section {Fence $\Fnc$}\label{s:CB}

Intuitively, the main function of $\Fnc$ is to prevent changes of
senior chains from rips. Only {\zlocks} and splits may change their
senior pointers, and thus their and their descendants' senior chains
(and heights).

Call $v$ \trm{hanging} if the $\zp l$-chain from $v$ has a long $\zp l$.\footnote
{
  An alternative more precise definition is possible: the $\zp
  l$-chain from $v$ to the long edge contains no splits with double
  children and no nodes that were ever {\zoff}.
}
An \trm {apex} is a {\zlow} $v$ with no {\zlow} children; 
 when $v$ becomes a switch it might split (or float).
An {\zon}-apex $v$ is \trm {loose} if it has no $\zp l$-children: it
can split and then merge prematurely (without completing a full
$\Fnc$ cycle, see below).
To assure ($\Fnc$.rip), $\Fnc$ needs to check that its tree has no
incident rips (including $\zp l$), 
but such a check is unreliable if a neighbor $v$ is
(1) hanging;
(2) childless {\zlow} with a hanging neighbor or a long edge;
(3) childless {\zlow} with a childless {\zlow} neighbor $u$ and a long
edge $uw$ to a {\zlow} $w$.\footnote
{
  In the last case, $u$ can split to $w$, while $w$ is {\zon}; changing
  $w$ to {\zoff} will result in $u$ prematurely merging into a double
  (changing its height and making $vu$ long); then $v$ can split to
  $u$ while $u$ is still {\zoff}, and have another neighbor split to $v$;
  then changing $u$ to {\zon} will result in the premature merge of $v$,
  changing its height.
}
Such $v$ can change height creating rips for its (possibly already
checked) neighbors.
So, in addition to rip-checking incident edges, $\Fnc$ must assure
that before getting an {\zoff} $\zp l$-ancestor,
(1) its {\zhi} neighbors will check that they are not hanging, and
(2) its childless {\zlow} neighbors $v$ will rip-check their edges and in
turn assure that their childless {\zlow} neighbors $u$ have no rips
$uw$ to a {\zlow} $w$. 
This requires two ``milestones'' in the {\zhi} nodes and three for the
childless {\zlow} nodes. So, next we describe the $\Fnc$ cycle which
achieves these ``milestones''; then we describe the rip-checking and
exiting from {\zlock}s.

\subsection {$\Fnc$ cycle}\label{s:CBcycle}
 $\Fnc$ cycle is initiated on a $\zp l$-tree from its root by switching
to {\zon} (``registering'' $\zp l$-pointers forming the tree; a $\zp
l$ pointer joining the tree after this registration will participate
only in the subsequent $\Fnc$ cycle).
 Unless specified otherwise, the parents and children below refer only
to these (registered $\zp l$-) tree edges.

{\bf Transitions.}
The $\Fnc$-cycle consists of two phases (0 and 1), each with three
states: {\start}, {\activ}, {\done}. 
 Intuitively, the goal of phase-1 is to provide assurance (to the
neighbors) of height preservation, while phase-0 is focused on
assuring no rips (for its own nodes).
In a regular $\Fnc$ cycle phase-0 is run once (following {\zoff} wave),
while phase-1 is potentially re-cycled repeatedly (until the next
{\zoff}), from an unregistered split.

In {\zhi} nodes the states function similarly to the classical children
game of fire-water-hay: with fire ({\start}, propagating up: from parent
to children) consuming hay ({\done}), but put out by water ({\activ},
propagating down: from children, when all {\activ}, to parent), which in
turn is absorbed by hay ({\done}, propagating down, similarly to {\activ}).

In {\zlow} nodes the transitions are slightly more complex: there
{\start}, {\done} and {\activ-0} propagate in the same  
directions as in {\zhi}, but {\activ-1} propagates from parent to
children.
More specifically, {\done-0} in {\zlow} nodes is delayed while
{\activ-0} (which enters a {\zlow} node only when all its {\zhi} children
enter {\done-0}) propagates to the root turning into {\activ-1} signal
propagating back towards {\done-0}. Then {\done-0} propagates on {\zlow}
(replacing {\activ-1}) towards root. Upon reaching the root, {\done}
changes into {\start-1}, which propagates up replacing {\done-0}.
Similar to phase-0, a {\zlow} {\start-1} does not change until its {\zhi}
children are all {\done-1}, but here it changes directly to {\done-1}
which proceeds towards the root (consuming {\start-1} parents). 
A root with all children {\done-1} changes to {\zoff}, signaling that
$\Fnc$ is finished on this tree.

If a node $v$ in {\done-1} (and all children in {\done-1}) splits or
uproots, then it recycles phase-1 on its subtree until changing to
{\zoff}: {\done-1} with low (also {\done-1}) parent changes to {start-1}.
 Thus, intuitively, 
{\start} propagates always from the parent to the children;
 and
{\done} --- from the children (when all are {\done}) to the parent; 
{\activ-0} propagates similarly to {\done}, while {\activ-1}---towards
border: as an echo (preceding {\done-1}) in {\zhi}, and as a signal in {\zlow}.

$\Fnc$ can mark nodes as {\zhi}, {\zlow}, apex and loose (in the draft and
certificate, see below), so that it is visible not only to the node
but also to its neighbors (loose, or even apex, status can be omitted,
then all apexes, or even all {\zlow}, would be treated as loose); the
algorithm description below uses this recorded {\zhi}/{\zlow} status.

{\bf Checks.}
$\Fnc$ needs to check that its tree has no incident rips, and that the
neighbors will not create them after the check is complete.
 {\Zlow} nodes ---unless loose--- need no such checks: they can change
neither senior chains nor heights until after the next change to {\zon}.
Thus, the following checks are performed:
In {\start-1}: a loose $u$ rip-checks all its edges before changing to {\done-1}.
In {\activ-1}: split $u$ rip-check its $\zp l$-pointer (delaying change
to {\done-1}). 
Also a loose {\activ-1} $w$ (which in {\zlow} occurs before {\start-1})
waits for each {\zlow} (loose) neighbor to be in phase-0 or
{\activ-1} before changing to {\done-0} (thus assuring correctness of
the {\start-1} check above). 
In {\activ-0}: a {\zhi} $v$ before changing to {\done-0}
  (1) rip-checks all edges, and 
  (2) waits for each 
    (a) {\zhi} neighbor $w$ to be in phase-0, or to enter {\activ-1}
and then enter {\start-1},
    (b) {\zlow} (loose) neighbor $w$ to be in {\start-0} or {\activ-0}
(assuring correctness of the subsequent {\activ-1} check above).
Finally, in {\start-0}, loose $v$ waits for the same events as in (2)
above before changing to {\activ-0}.

{\bf Splits: borrowing a pointer.} 
The above checking requires a pointer to ``rotate'' over the
node's neighbors. This (soft) pointer can use the unused
hard pointer in the singles or doubles. In splits no spare hard
pointer is available, however (instead of adding a hard pointer) we
can ``borrow'' a pointer from the $\zp b$-parent as follows. When a
split $w$ needs to use an extra pointer, $w$ requests help from its
$\zp b$-parent ({\zlow}, and thus always single) $v$. Such $v$ goes
around pointing at its needy $\zp b$-children with the ``lending''
pointer. Such a ``lending'' pointer on $w$ (there can be at most one),
can implement its $\zp b$ pointer (in the opposite direction),
allowing $w$ to use the corresponding hard pointer for other
purposes. When $w$ is done using its client pointer, it can free the
``lending'' pointer, allowing $v$ to lend it to its other $\zp
b$-children.
 Each split needs to borrow a pointer only when in
{\activ-0}, so it can request help from its $\zp b$-parent at most
once in a $\Fnc$ cycle, and thus at most two times total before it
merges.
Since split $w$ might be waiting for its {\zlow} (loose) neighbors to be
in {\start-0} or {\activ-0}, the lending {\zlow} $v$ should do the
lending in the same states (otherwise, a deadlock can occur).

{\bf Rip-checking} is more efficient if it runs on small
groups, called \trm {clients}.
The client tree is formed of the registered $\zp l$ when the $\Fnc$
tree is formed. The subsequent change of the tree to {\zoff} changes the
clients into \trm {servers}, functioning in a similar fashion (the
{\zoff} may lead to new splits, so the servers are along senior
pointer trees).
The rip-checking is implemented by interactions of clients and servers
as described below.
 Each client must be large enough to contain
its own height ({\zrise} from the root) $\rho$; for
$\rho=O(1)$ the client is just one node, making its rips instantly detectable.
In fact, each client should contain $\theta(\lg\rho)$ nodes and is
computed (allocated and initialized) from the parent client.\footnote
 {For example, let $i$ be the smallest such that the subtree $T_v(i)$
 of all descendants of $v$ at distance $\le i$ from $v$ contains
 $|T_v(i)|\ge \lg\rho$ nodes. Then $T_v(i)$ forms a client of $v$ if
 $|T_v(i)|<2\lg\rho$. Otherwise, additional clients are formed (\eg
 from the leaves of $T_v(i)$). These additional clients might not be
 able to form a separate connected subtree, but their nodes can still
 communicate (as in sec.~\ref{s:CA-TM}) through the nodes of the
 parent client (thus nodes might need additional child support fields).
 Finally, subtrees of the nodes which are too small to have clients of
 their own join the parent, possibly splitting it into more clients
 similarly to the above. The child support does not introduce any
 overhead, since similar communication needs to be provided, whether
 for the own or the child client.}
 Each client also computes a timer (as in sec.~\ref{s:pp}) which
re-checks repeatedly both the client size (compared to its {\zrise}
$\rho$, which in turn is checked with the parent client) and the
upper bound on its computation time (wlog, assume it is $2^t-1$ for some
$t$; then co-located step counters are trivially assured never to
exceed it).
 
To detect rips, each client is first re-initialized (to assure that it
is not created by the adversary) and then  goes through its
edges one at a time, using a special client pointer, attaching it as a
leaf to the server.
 Each server periodically registers the attached client pointers, then
verifies its correctness (from the root), and then
serves its height to all the registered clients one bit at a time (the
clients that attached to the server after its registration stage are
ignored by the server until the next registration). Each client, upon
receiving this height, compares it with its own height value. The
client-server interface is across the (client pointer) edge connecting
them and can work as follows:
 Let the server height be encoded in ternary, so that no two
consequent digits are the same (\eg we can use ``2'' as a separator
between 0 and 1 digits; more efficiently, to encode the next bit use the
two values different from the current one: the greater to encode 1, and
the smaller for 0).\footnote
{ A client not copying the served bit delays the step in its server
 parent node (\ie its $\bmod3$ counter is not incremented). Similarly,
 the server not serving the next bit after the current one is copied
 delays all its clients' clocks.
 Thus a client might indirectly delay a different client of the same
 server. However, since each client has only one server parent, after
 a server serves a bit, all clients independently and in parallel
 must consume it promptly, thus avoiding deadlocks.
 After $\Si$ stabilization, such delays are $O(\Delta \lg d)$; and
 before it, they do not impact any commitments.
}
 The step counters and the timer assure that even the adversarially
initiated clients and servers terminate promptly ($\Delta (\lg
d)^{O(1)}$ after $\Si$ stabilization\footnote
 {Indeed, if for the client (the same for servers) its $\Delta=O(\lg
 d)$ then the $\Delta$
 factor can be ignored; otherwise, if $v$ has $>2\lg d$ children then
 these children (without grand-children of $v$!) form one or more
 clients of $<2\lg d$ nodes, whose communication has a $\Delta$
 delay due to the information going through $v$, so any polynomial
 algorithm can be executed by the client in $\Delta(\lg d)^{O(1)}$.
}).
If a rip is detected then this and the neighboring trees need to be
restructured, so we change the rip servers to {\em void} to
initiate the following restarting procedure, used also in the case
of {\zcrash}ing.

\subsection {Restarting}\label{s:redrw}
A {\zcrash} might corrupt computations in the clients and servers, so
it is safer to reconstruct them, \eg as follows. Let $\Fnc$ keep a
special {\reborn} flag, typically set to false, but with the default value
true. So, when the node is {\zcrash}ed (\inc into a root) and then
opened by $\Si$, it is still {\reborn}.
Servers adjacent to a {\reborn} are marked as {\em void}
(starting from the {\reborn}'s neighbor and spreading through
the whole server tree); cleared server fields in nodes that were
{\zcrash}ed (and exited) are also interpreted as {\em void}.
Both {\zhi} and {\zlow} {\start-0} (propagating along the {\zon} wave)
freezes at the $\zp l$-pointer of a split with a void server,
neither crossing the pointer nor changing till the server changes
to non-void.
If $v$ is adjacent to a void certificate, then $v$'s client-tree (if any)
is cleared: $v$'s {\em void-client} propagates from client-child to its
parent until reaching the client's root
there the client
is cleared, causing the descendant clients to clear as well (the void
server's origin also clears its client).
 If the {\em void-client} mark (on its way to the root) meets an
{\zoff} wave moving this client to server fields, then the move
leaves the resulting server void (since it
was just moved from the client fields, this new server
does not intersect any clients, so this process does not propagate any
further).
A {\reborn} flag is cleared when all adjacent servers and clients are void.

When a void server tree has no clients in any of its nodes and no
adjacent {\reborn}, $\Fnc$ computes {\em re-clients} on the void
server tree (similar to clients, but not on $\zp l$-tree).
A re-client near a {\reborn} is cleared similarly
to the client (the {\reborn} could have been the re-client's child
potentially corrupting it): it changes to void re-draft
which propagates to the re-client tree root and is erased from there.

When a re-client is constructed, it checks (as part of an echo state
propagating from re-client tree leaves to roots, when a node's children are
all in echo) that neither {\reborn} nor clients are adjacent; then
re-clients are copied to servers (non-void;  possibly
changing the sign of $\zh{=}\pm0$ at root child accordingly) from the
root up the server tree.

\subsection {$\Fnc$ Performance}

In this section all the distances are along the tree edges described
in the previous section, and we assume that $\Si$ has stabilized.

A {\zhi} {\start-1} changes to {\activ-1} within $O(d)$. 
Indeed, a {\zhi} {\done}  with {\start-1} parent changes to {\start-1}
within a step.
So, the distance from a {\zhi} $v$ in {\start-1} to the nearest
{\done-1} descendant as above (\ie with no {\activ} in between) grows
each step till (within $O(d)$) none remain (only {\start} can be a
parent of start; similarly, {\done} can have only {\done} children).
 a {\zhi} {\start-1} with neither {\done} nor {\start-1} children (\ie
only {\activ-1}, if any) changes to {\activ-1}, so the distance to the
furthest {\zhi} {\start-1} descendant decreases each step and any {\zhi}
{\start-1} changes to {\activ-1} within $O(d)$.

a {\zhi} {\activ-1} changes to {\done-1} (or {\zoff}) within $O(d)+\Delta
(\lg d)^{O(1)}$. 
Indeed, each {\activ-1} split must rip-check its $\zp l$ (which takes
$\Delta (\lg d)^{O(1)}$ steps), after which
each {\zhi} {\activ-1} with all children (if any) in {\done-1}
changes to {\done-1} within a step, (unless its parent
is {\zoff}).

A loose {\start-1} $v$ with {\done-1} children checks the lengths of
all its edges within $\Delta^2 (\lg d)^{O(1)}$ steps: each edge is
rip-checked in  $\Delta^2 (\lg d)^{O(1)}$ and a client can have
$O(\Delta \lg d)$ edges, checked one at a time.
Once this rip-check is completed, $v$ changes to {\done-1}.

If a {\zhi} $v$ in {\done-1} has a split ancestor with
unregistered $\zp l$, then it too changes to {\done-1} within
$O(d)+\Delta (\lg d)^{O(1)}$ and then changes to {\start-1} or {\zoff};
and then in $O(d)$ steps more $v$ changes to {\start-1}, or {\zoff} (and
then to {\start-0}) as well.  Thus any {\zhi} $v$ enters {\activ-1} and
then {\start-1} (or changes to {\start-0}).  Similarly, a {\zlow} $v$ in
{\start-1} or {\done-1} changes to {\start-0}, but with the additional
$\Delta^2 (\lg d)^{O(1)}$ delay due to the loose nodes.

Let $t_{l1}\edf d + \Delta^2(\lg d)^{O(1)}$ be the time required by a
loose $w$ to be seen in phase-0 or {\activ-1}.
Let $t_{u1}\edf d + \Delta^(\lg d)^{O(1)}$ be the time required by an
{\zhi} $w$ to be seen in phase-0, or to enter {\activ-1} and then {\start-1}.

{\Zlow} {\activ-1}, {\done-0} change to {\start-0} within $O(\Delta t_{l1})$.
Indeed, within $O(d)$ {\zlow} {\activ-1} has no {\activ-0} descendants:
the closest of these changes to {\activ-1} in one step.
A loose {\activ-1} changes to {\done-0} within $\Delta t_{l1}$: after
waiting for each {\zlow} (loose) neighbor to be in phase-0 or
{\activ-1}.
A non-loose {\zlow} {\activ-1} with only {\done-0} children changes to {\done-0}
in a step, and so the distance to the farthest {\activ-1} decreases.
A root with only {\done-0} children changes to {\start-1}, which
changes to {\start-0}, since it has a {\zlow} descendant, which will
change to {\start-0} too $O(d)$ steps later.

a {\zhi} {\start-0} changes to {\activ-0} in $O(d)$.
Indeed, any {\start-0} has no {\zoff} descendants within $O(d)$. 
Then a {\zhi} {\start-0} with no {\start-0} children (all, if any, are
{\activ-0}) changes to {\activ-0}, 
so the distance to the furthest {\zhi} {\start-0} descendant decreases
each step.

Before a {\zhi} {\activ-0} can change to {\done-0} and a loose
{\start-0} to {\activ-0}, the rip-checks for the {\zhi} and
neighbor state checks for both {\zhi} and loose need to be performed. For
{\zhi}, these checks can be done by all the nodes in parallel.
Each client needs to check $O(\Delta \lg d)$ edges, each edge checking
taking $\Delta (\lg d)^{O(1)}$ steps (plus a delay due to splits
borrowing pointers).

In addition to rip-checking, {\zhi} {\activ-0} and loose {\start-0}
wait $\Delta t_{l1}>t_{u1}$ to see each {\zlow} (loose) neighbor in
{\start-0} or {\activ-0} 
(this dominates the check of the {\zhi} neighbors, which still needs to
be performed). Both of these {\activ-0} checks can be done by all {\zhi}
in parallel (with the client restrictions for the rip-check) and
both requires pointers (thus splits still need to borrow them from
their $\zp b$-parents). The checking of the states dominates the
rip-checking, so the time it takes a {\zhi} {\activ-0} $v$ to check
all of its edges is $O(\Delta^2 t_{l1})$. Thus, a split may need to wait
for $t_{lend}\edf O(\Delta^3 t_{l1})$ steps before its $\zp b$-parent
could lend it the pointer. Thus, all {\zhi} {\activ-0} $v$ will
all complete their checking within $O(\Delta^3 t_{l1})$ and then any
{\zhi} {\activ-0} with no {\activ-0} children will change to {\done-0}.
So, within $O(\Delta^3 t_{l1})$ steps ($O(d)$ time for {\done-0}
propagation is absorbed since $d=O(t_{l1})$) all {\zhi} {\start-0} change to
{\done-0}.

A loose {\start-0} does not need to borrow a pointer, and so exits to {\activ-0}
within $\Delta^2 t_{l1}$. The propagation of {\activ-0},
{\activ-1} and {\done-0} in both directions on the ancestors of loose
$v$ takes
additional $O(d)$ (absorbed in the asymptotics of $t_{l1}$). Thus all
{\start-0} change to {\done-0} within $t_0\edf O(\Delta^3 t_{l1})$, which
also provides the asymptotic upper bound on the $\Fnc$ cycle time: the
time within which $\Fnc$ turns {\zoff} at a root (fulfilling
($\Fnc$.{{\zoff}})).

\subsection {$\Fnc$ Correctness}
Assuring ($\Fnc$.{{\zoff}}) is demonstrated above.

Any senior chain contains at most one $\zp b$. 
Indeed, a split-$\zp b$ separates {\zhi} nodes from {\zlow} ones,
and chains from {\zlow} nodes can (legally) contain only {\zlow} (or
{\zlock}).

A node with {\zoff} descendants can only be in {\start-0} or
{\zoff}, together with the above assuring
($\Fnc$.{cln}).

A {\zcrash} of $v$ marks it {\reborn}, which voids the server trees of $v$
and its neighbors, and clears the client trees adjacent to these void
trees. This effectively freezes $\Fnc$ in the respective nodes. Then
{\reborn} it reset to false, and void servers as well as cleared
clients are recomputed.
Thus, the tree of $v$ and the adjacent trees have new (uncorrupted by
{\zcrash}) servers; the client trees of $v$ and its distance two
neighbors are also recomputed and restart their $\Fnc$ cycles (and will
not let $\Fnc$ turn {\zoff} when detecting a long edge). Thus, this
situation essentially as if the leaves of each of these trees have
just changed from {\zoff} to {\zon} (binding corresponding edges), and so it
is now reduced to the following.

Assume now no {\zcrash}es taking place.
Consider $v$ changing its senior chain while $vw$ is a rip.
 Then $v$ is either {\zhi} or loose: an apex can split,
but ---unless loose--- will go through another $\Fnc$ cycle before
merging (and thus changing its senior chain).
Consider the interval from the last moment $v$ was {\start-0} with an
{\zoff} descendant  (there was one that made $vw$ bound) and until $\Fnc$
turns {\zoff} before the senior chain change. 

$\Fnc$ rip-checks all edges incident to {\zhi} and loose nodes of the
tree ({\start-0} guarantees correctness).
Thus, during the rip-check, $vw$ was not long, so $w$ must have
changed its height after the rip-check.

If $w$ is {\zhi}, then $v$ observes it in phase-0, therefore
ancestors will rip-check their $\zp l$ before $\Fnc$ turns {\zoff} at the
root (and so before merging). Thus, {\zhi} $w$ cannot create the rip.

A {\zlow} $w$ cannot change height unless it is loose. Then $v$ had to
wait for $w$ to be in {\start-0} or {\activ-0}.
A loose $w$ can change height only if it splits and then merges
prematurely:
(i) with the new parent $u$ which was {\zon} during the split of $w$, then
$w$ merges (possibly without any $\Fnc$ checks) when changing to {\zoff};
(ii) with the new parent $u$ which was $\znew$ during the split of $w$,
then before $w$ changes to {\zon}, some splits pointed at it and $u$
remained non-single, so $w$ merges when changing to {\zon}.
Before $w$ splits, it rip-checks $wu$, so if $w$ changes height then
$u$ must changes height after the check and before $w$ merges.
In case (i) this possibility is eliminated by $w$ waiting (in
{\activ-1}) for $u$ to be in phase-0 or {\activ-1}. Then $u$
rip-checks its $\zp l$-chain if {\zhi}; if {\zlow}, $u$ cannot change
height either: even if it splits $u$ cannot merge when changing to {\zoff}
(since it has children), and so rip-check of $w$ prevents its change
of height.
In case (ii) $u$ rip-checks its edges before splitting; if its new parent
change height after the check, $u$ would merge prematurely into
single, and $w$ would not merge prematurely. Thus $w$ cannot change height.

Therefore, $\Fnc$ assures ($\Fnc$.{rip}).

Finally, it remains to satisfy ($\Fnc$.{sgn}). This is
done by the clients computing $\lnd((h(v)+1)/3)$ in addition to $h(v)$
for each node to be used in case it floats to $\zh=0$.

\appendix \section*{APPENDICES}

\section {Sketch for $\Si$}

$\Si$ controls {\zcrash}ed roots (since $\Si$ is invoked last, it can
{\zcrash} them back if the roots are uprooted by other protocols) and
{\zlocks}, keeping its own pointers in them. Intuitively, these
pointers must always point down, according to the $\Si$ own notion of
height; the {\zlock} ($\Si$ pointer) cycles are broken with the 
help of {\em acyclicity certificates} (similar to those of
\cite{IL92}) maintained in the {\zlock} pointer chains.
 $\Si$ {\zcrash}es its long edges; changing the pointers and requiring
adjustment of the certificates. 
Unlike the clients and servers of $\Fnc$, these certificates must be
adjusted locally (on a sufficiently small interval of the certificate:
the whole certificate tree is too big).
Furthermore, we will define the long edges in such a way that if a
configuration has no {\zstub}s, it will be guaranteed to have long
edges, which can be promptly detected and {\zcrash}ed.

Thus we will reduce $\Si$ to 
(1) $\AM$: {\zlock} cycle Cutter, and
(2) $\bfs$: Dropper; their performance parameters
$\tcc,\tcb;\tbfs$ are functions of $d,\Delta,n$ and sometimes
other aspects of the configuration.

\subsection{Reduction}

\paragraph{Interface.} {\bf Fields:}
 {\AM, \bfs} share $\zprc, \zprb$ in each {\zlock} ($\bvec{v}\,\edf
 v.\zprB\edf v.\zprb$ if ${\ne}v$, else $v.\zprc$;  $v$ is a \trm{\zgrnd} if
$\bvec{v}{=}v$; $\dbfs$ is the length of the longest $\zprB$-chain).
 An additional bit $\zbl$ indicates {\em long} $\zprB$ (used
mainly for the contracts). 
  
{\bf Automatic (local) actions:}
A {\zlock} $v$ adjacent to a {\zgrnd}${\ne}v$ is {\zcrash}ed if $v$
is {\zgrnd}, or $\bvec{v}$ is not a {\zgrnd}, or
$v{\ne}v.\zprb{\ne}v.\zprc{\ne}v$. 
{\Zcrash} always loops $\zprc$, and sets $\zprb$
to an adjacent {\zgrnd} (possibly resulting from an open root) if
there is one; if not, $\zprb$ is looped too (we call such {\zcrash} \trm{\grnd}), except $\bfs$ can also
set $\zprb$ to an adjacent {\zlock} with non-loop $\zprc$.
 (So, after the first step, {\zgrnd} nodes are never adjacent; and for {\zlock}
$v{=}v.\zprc$ either $u{=}v.\zprb$ is a {\zgrnd} or $u.\zprc{\ne}u$).
A {\zlock} $v$ decrements $\zh$ (whenever allowed by the interface of
Sec.~\ref{s:prot}) if $v$ is {\zgrnd} with $v.\zh{\ne}{-}1$, else if
$v.\zh{\not\equiv}(\bvec{v}).\zh{+}1\!\!{\pmod3}$. 
A {\zlock} $v$ sets $v.\zbl{\gets}1$ if $(\bvec{v}).\zbl{=}1$.

{\bf Permissions:}
$\AM$ is invoked in (and reads fields of) only {\zlock}s; $\bfs$ acts in all
$v$.
$\AM,\bfs$ can {\zcrash} any node.
$\bfs$ can also set $v.\zbl{\gets}1$ of any {\zlock} $v$. 
When $\zbl{=}1$ for $v$ and all its {\zlock} $\zprB$-children, $\bfs$
can change $\zprb$ to an adjacent lock $u$ with non-looping $\zprc$
and $u.\zbl{=}0$, resetting $v.\zbl{\gets}0$. 
$\bfs$ can loop $\zprb$, when $\zprb{=}\zprc$ and $\zbl{=}0$.
$\AM$ can set $\zprc{\gets}\zprb$ for any {\zlock} $v$. 
$\bfs$ can also change the sign of $\zh{=}\pm0$ in {\zlock}s, and open
{\zon} {\zlock}s by swapping $\zP{l}v,\zP{b}v$ (both while obeying Interface
permissions of Sec.~\ref{s:prot}). 

\paragraph{Height.}
 First, let $v$ be a {\zlock}. Then  $\hsi(v){\edf}{-}1$ if $v$ is {\zgrnd},
else $\hsi(v){\edf}\hsi(\bvec{v}){+}1$ unless $v.\zbl{=}1$ ---
in this case $\hsi(v)$ is unchanged from its previous value (undefined
before the first action). 

Now, let $v$ be open.
 Then define $\hsii{i}(v){\edf}h{\in}[{-}1,3\cdot2^i{-}1)$, for unique $h$
such that  ${w.\zh\equiv h{+}\rho_{v,w}}\!\pmod{3}$
for all $w$ on some (sufficiently long: $O(2^i)$) open
$\zp{B}$-chain from $v$, where $\rho_{v,w}$ is the chain {\zrise} from
 $v$ to $w$, 
 and if $w.\zh{=}\pm0$ then its sign is $\lnd((h{+}\rho_{v,w})/3)$, if $w$
is ground then $h{+}\rho_{v,w}{=}w.\zh$.
 If no $h'\ge3\cdot2^i{-}1$ satisfies the same condition on the same
chain (intuitively, when the $O(2^i)$ chain contains ground or two
marks with non-0 {\zrise} between them), then  we say that
$\hsii{i}(v)$ is \trm{final} and write $\hsi(v){=}h$. If
$\hsii{i}(v)$ is defined but not final, we say $\hsi(v)\ge3\cdot2^i{-}1$. 
 If more than one $h{\in}[{-}1,3\cdot2^i{-}1)$ satisfies the above
condition for the maximal open $\zp{B}$-chain (the chain is too short,
anchored in a {\zlock}), then $\hsii{i}(v){\edf}*$, and $\hsi(v)$ is
unchanged from its previous value. 
 If not even one such $h$ exists (signs of $\zh{=}{\pm}0$ are
inconsistent with $\lnd$), then $\hsii{i}(v){\edf}\bot$. 

\paragraph {$i$-rips. }
 An edge $vu$ is an \trm{$i$-rip} if 
(a) $v,u$ are open,
 $\hsii{i}(v){-}\hsii{i}(u){\not\equiv}0,\pm1\!\!\!\!{\pmod{3{\cdot}2^i}}$, 
or $\hsii{i}(v){=}\bot$; or
(b) $v$ is a {\zlock} with $\hsi(v){<}3{\cdot2}^i{-}1$ 
and $\hsi(u){>}\hsi(v){+}1$.
 The $i$-rip $vu$ is \trm{fixed} 
when $u$ is a {\zlock} and $\hsi(u){\le}\hsi(v){+}1$.
$v$ \trm{matures} 
when ground or {\zgrnd}, 
when resets $v.\zbl{\gets}0$, and after $\tbfs(\hsi(v))$ steps.

\paragraph{ {\bfs} commitments:}
(1) In mature $v$, $\bfs$ 
 (a) can  reset $v.\zbl{\gets}0$ (and change $v.\zprB$) only if
decreasing $\hsi(v)$; 
 (b)~can open $v$ only with no $i$-rips, but (c) cannot {\grnd} $v$.
(2) $\bfs$ fixes $i$-rip within $\tbfs(2^i)$ (${>}\tcb(2^i)$ below).
(3)~If orientation remains flat with all non-{\zgrnd} {\zlock}
   pointers down, then {\bfs} promptly opens {\zlock}s.

\paragraph{ $\AM$ commitments:}
(1) After the initial $\tcc$ steps, $\AM$ assures a
    {\zgrnd} if there are {\zlock}s.
(2) $\AM$ un-loops $v.\zprc$ in non-{\zgrnd} {\zlock} $v$ within
    $\tcb(\hsi(v))$.
(3) $\AM$ does not {\zcrash} $\tcb(\dbfs)$. 
(4) $\AM$ merges $v.\zprc\gets v.\zprB$ for every {\zlock} $v$ within
    $\tcb(\dbfs)$. 

\subsection{Correctness}

\BCl
(\bfs.2) promptly assures {\zstub}s.
\ECl
This follows directly from the fact that
{\em any configuration with no {\zstub}s contains a $\ceil{\lg{(d{+}1)}}$-rip.}

Indeed, set $k{=}\ceil{\lg{(d{+}1)}}$ and let there be no {\zstub}s. 
Then there is $\zp l$-cycle; by ($\Fnc$.cln) it is all one phase,
thus its $\Se$ pointers do not change.
 By ($\Le$.ht), it must also contain a $\zp l$-chain from $v$ to
$w$ of {\zrise} $d{+}1$. 
If 
$\hsii{k}(w){\not\equiv}\hsii{k}(v){+}d{+}1 \!{\pmod{3{\cdot}2^{k}}}$,
then some $\zp l$ in the chain is a $k$-rip.
 Else, consider a shortest path  $v_0...v_s, v_0{=}v,$ $v_s{=}w,$ $s{\le}d$.
Since $s{<}d{+}1{<}3(d{+}1){-}s$, for at least one $j{<}s$ the edge
$v_jv_{j+1}$ is a $k$-rip. \qed

\BCl\label{cl:grnd}
(\bfs.2) and (\AM.1) assure {\zgrnd} or ground any time after a
prompt initial period.
\ECl
Indeed, assuming $\tcc,\tbfs({\le}2d)$ are prompt, (\bfs.2) promptly
assures a root or {\zgrnd} if there were no {\zlock}s initially; otherwise,
(\AM.1) promptly assures {\zgrnd}.
 A {\zgrnd} may change only to a root. 
A root $r$ may uproot; then its $\zp l$-chain leads either to another root,
or {\zlock} (then {\zgrnd} is assured by $\AM$), or cycle. By
($\Le$.ht) the cycle in the last case must be unbalanced, which
implies that $v$ was not bound ($\Fnc$.rip) and remains ground (since
the cycle contains only $\bFnc{=}1$ nodes by ($\Fnc$.cln)).
 Furthermore, if there are no more {\zstub}s, there must be a
 $\ceil{\lg{(d{+}1)}}$-rip, which was there even before the uprooting.
\qed

For the next claim let us measure time as the number of activations
(of any nodes), starting from some initial configuration at time
denoted as $0$. 
Let $h_t(v)$ be $\hsi(v)$ at time $t$.
We say that node $v$ has $(m,h,t)$-trajectory if in the $0$ to $t$
period (inclusively) the minimum height $\hsi(v)$ of $v$ when mature
is $m$, and at the end of this period $h_t(v)=h$. 

\BCl If $v$ has $(m,h,t)$-trajectory and $h>m+2$ then for any neighbor
$w\in \E(v)$ there are $t'<t$, $m',h'$, such that $w$ has
$(m',h',t')$-trajectory and $|m-m'| \le 2, |h-h'| \le 1$. \ECl

{\em Proof:} Let $v$ have $(m,h,t)$-trajectory and $h>m+2$. Let $t'$ be
the largest such that $h_{t'+1}(v)=h_{t'}(v)+1=h$ (\ie it is the last
float to $h$ of the trajectory of $v$). Then $v$ has
$(m,h,t'+1)$-trajectory.

Suppose that the $(m',h',t')$-trajectory of $w$ violates either $|m-m'|
\le 2$ or $|h-h'| \le 1$. Consider the (first) time $i$ when $v$ is at
the minimum height $m=h_i(v)$ and floats at the next step $h_{i+1}(v)=m+
1$. 
(Mature $v$ cannot increase $\hsi(v)$, other than by floating
($\bfs$.1); only the first float may be adjacent to rips ($\Fnc$.rip).)
Since $h> m+1$, $v$ must float again, now to height $m+2$. At that time,
$\hsi(w)$ will be defined (and ${=}h(w)$) and will have the value $m+1$ or
$m+2$. Thus, $m'\le m+2$. Similar argument provides $m\le m'+2$,
showing $|m-m'| \le 2$.

The above implies that at time $t'$ both $h_{t'}(v)$ and $h_{t'}(w)$ are
defined. Furthermore, to permit floating of $v$, we must have
$h_{t'}(w)$ be either $h-1$ or $h$. \qed

\BC\label{cor:low}
 If $v$ rises by $d{+}1$ while remaining at $\hsi(v){>}2d$ then
 during that period $\hsi(u){>}0$ for all $u$.
\EC
Proof by induction on distance $k$ from $v$ to (any) $u$ (and using Claim
for the inductive step).

\BC
 If $v$ is a ground or {\zgrnd}, then $\hsi(v)$ remains $O(d)$.
\EC
 This corollary follows from the previous and Claim~\ref{cl:grnd} ($v$
 is mature after 1 step).

\BCl
Given $v, \hsi(v){=}O(d)$, (\bfs.1) promptly assures $\hsi(u)=O(d)$ for
all $u$.
\ECl

Assume $\tbfs(h),\tcb(h)$ are polynomial in $h$.
Let $v{=}u_0u_1{\ldots}u_k{=}u$ be the shortest path from $v$ to $u$,
and let $\hsi(v){\le}h=O(d)$.
 Then if $\hsi(u_i){\le}h{+}i$ then within $O(\tcb(h{+}i))$ $v$ is
open or has a non-loop $\zprc$ (\AM.2), and within $O(\tbfs(h{+}i))$
more ($\bfs$.2) assures $\hsi(u_{i+1}){\le}h{+}i{+}1$. \qed

\BCl
$\AM$ and $\bfs$ both promptly stop {\grnd}ing.
\ECl

The previous claim implies that all $v$ promptly mature and $\dbfs$ is
promptly $O(d)$. Then (\bfs.1c) stops $\bfs$ {\grnd}ing, and (\AM.3)
promptly stops $\AM$ {\grnd}ing. \qed

\BCl
$i$-rips disappear promptly after {\grnd}ing stops.
\ECl
The minimum $\hsi(v)$ with $i$-rip $vu$ increases by (\bfs.2) within
$\tbfs(2^i)$. \qed

\BL
$\bfs$ (and $\Si$) promptly stabilize.
\EL

After there remains no $i$-rips for any $i$ (see previous two claims),
$\zprc$ are promptly merged into non-loop $\zprb$, so non-{\zgrnd}
{\zlock}s $\zprc$ point down. Then, (\bfs.3) assures that {\zlock}s
are opened, stabilizing $\Si$.
\qed

\subsection{$\AM$ sketch}
$\AM$ consists of two protocols {\em Checker} $\CC$ and {\em Mender}
$\CB$, both sharing acyclicity certificate in special {\zlock} fields.
Intuitively, $\CC$ checks certificate {\zcrash}ing $\zprc$ cycles.
$\CC$ can also check certificate drafts along $\zprB$-chains to avoid
delayed {\zcrash}es when the drafts are moved to the official
certificates along the (possibly merged) $\zprc$-chains.  $\CB$ mends
the certificates when $\zprc$-chains change, and extends them to new
{\zlock}s.
So, $\CC$ write access is limited only to {\zcrash}. $\CB$ reads and
writes certificate fields in {\zlock}s, merges $\zprc{\gets}\zprb$
$\CC$ promptly (in $\tcc$) breaks any $\zprc$-cycle, thus assuring
(\AM.1). $\CC$ can verify the correctness of certificate on an
$k$-long chain in poly($k$) time, allowing to assure (\bfs.3).
$\CB$ assures that its modification to the certificates will not
harm their correctness (so only ill-initialized certificates and/or
processes can cause $\CC$ to {\zcrash} the certificates). When all the
certificate chains are short, the certificates can be verified and the
$\CC$ {\zcrash}es stop.

$\CC$ can use the acyclicity certificates similar to those in
\cite{IL92} (see below).
 {Unlike the certificates of $\Fnc$, the acyclicity certificates here
 cannot be reconstructed on the whole tree (as it might be too deep)
 and so they must be adjusted locally.
 When one of the endpoints is {\zopened}, the adjustment is simple: the
 {\zopened} node is either {\zcrash}ed into root or the certificate is extended
 just by one --- trivial for many certificates.}

\subsubsection {Acyclicity Certificates}
\label{s:certs}\label{s:thue}

We illustrate the idea of acyclicity certificates, by briefly
sketching a variant used in \cite{IL92}. While there certificate 
was constructed along the dfs traversal path of a tree, here we define
using tree height.

Define $\mu(k)=-0$ iff $\sum_i k_i$ is odd and $>1$; $\mu(k)=+0$
otherwise.\footnote
{
  This is a variant of {\em Thue} (or {\em Thue-Morse}) sequence
  \cite{Thue-12} defined as $\theta(k)\edf\sum_i k_i\bmod{2}$, where
  $k_i$ is the $i$-th bit of $k$.
}
 In section~\ref{s:ht} we defined a similar sequence $\lnd$. Either of
 these two (and possibly some others) can be used to break symmetry:
 We say string $x=x_1x_2\ldots x_k$ is {\em asymmetric}
if it has one or two (separated by a special mark) segments of $\mu$
or $\lnd$ embedded in its digits (one sequence bit per constant number
of string digits).
  For simplicity, we ignore other ways to break symmetry. 
  Asymmetry is required for organizing (hierarchical) computations 
 (and for this reason $\lnd(h(v)/3)$ is made available to $\Si$,
  $\bfs$ specifically, via $\zh=\pm0$). 

Let us cut off the tail of each binary string $k$ according to some
rule, say, the shortest one starting with $00$ (assume binary
representation of any $k$ starts with $00$). Let us fix a natural
representation of all integers $j>2$ by such tails $\jj$ and call $j$
the {\em suffix} $\s(k)$ of $k$. For a string $\x$, define $\r(\x,k)$ to
be $\x_{\s(k)}$ if $\s(k)\le\|\x\|$, and special symbol $\zund$
otherwise. Then $\a[k]=\r(k,k)$, and $\a(k)=\langle\a[k],\mu(k)\rangle$.
\footnote
 {Inclusion of $\mu$ in $\a$ makes it asymmetric but otherwise is useful
only for $<\! 40$-bit segments. Also, $\mu(k)$ could be used instead of
$\zund$ if $i>\|k\|$ in $\a[k]$, but this complicates the coding and
thus is skipped. It is also possible to reformulate the definition
using $\lnd$ instead of $\mu$.}
 Let ${\cal L}_{\a}$ be the set of all segments of $\a$. ${\cal L}_{\a}$ can be
recognized in polynomial time.

\BL\label{l:alpha} Any string of the form $ss$, $\|s\|\!>\!2$, contains
segment $y\!\not\in\! {\cal L}_{\a}$, $\|y\|\!=\!(\log\|s\|)^2\!+\!o(1)$. \EL

Other variants of $\a$ can be devised to provide greater efficiency or
other desirable properties (\eg one such variant was proposed in
\cite{IL92}).

For a language ${\cal L}$ of strings define a $\T({\cal L})$ to be the language of trees,
such that any root-leaf path contains a string in ${\cal L}$, and any equal
length strings on down-paths ending at the same node are identical.

Let $T_A(X_T)$ be a tree $T$ of cellular automata $A$ starting in the
initial state with unchanging input $X_T$. We say that $T_A(X_T)$ {\em
rejects} $X_T$ if some of the automata enter a {\em reject} state.
 Language $\T$ of trees is {\em $t$-recognized} by $A$ if for all $T$,
 $T_A(X_T)$ (1) rejects within $t(k)$ steps those $X_T$,
which contain a subtree $Y\not\in \T$ of depth $k$; and (2) reject none
of the $X$ with all subtrees in $\T$.
 For asynchronous self-stabilizing automata, requirement (1) extends to
arbitrary starting configurations and to trees rooted in a cycle;
requirement (2) extends to the case when ancestors or children
branches of the tree are cut off during the computation.

\BL\label{l:ca-det} For any polynomial time language ${\cal L}$ of asymmetric
strings, $\T({\cal L})$ is recognizable in polynomial time by self-stabilizing
protocols on asynchronous cellular tree-automata. \EL

\subsection{{\bfs} sketch}

$\bfs$ maintains groups somewhat similar to servers and clients of $\Fnc$. 
Each group maintains a contiguous segment of an asymmetric sequence
(\eg $\mu$ or $\lnd$ above) and contains the height of (or a lower bound,
if near a sufficiently low group). 
This allows $\bfs$ to hierarchically check for $i$-rips using the same
mechanisms as the acyclicity certificates above. Intuitively, a group,
working as a client, checks each of its incident edges one at a time
(non-hierarchically, since we are interested only in the groups at
$O(d)$ height). However, the servers need to be organized
hierarchically, storing also the pointer address in the hierarchical
sub-groups to the edges being served. Then even a large group can quickly
detect a low adjacent group. For rips with sufficiently large height
difference, the subgroup of the appropriate hierarchy level changes
the tree as a unit. This may break the original group, but the
remaining contiguous segments of asymmetric
strings will be sufficiently large to support the subgroups with the
sufficiently large lower bounds on height (sufficiently larger than
the defecting subgroup's new height).

$\bfs$ extends its the above data structures to the open trees rooted
in {\zlock}s. There, it computes the height using $\lnd$ embedded in
$\zh{=}\pm0$. If the open tree is not large enough (does not contain
two marks with non-0 {\zrise} between them), nor contains height
information written there by $\bfs$, then $\bfs$ {\zcrash}es the whole
tree.
$\bfs$ treats open {\zlow} and {\zhi} branches separately: the {\zlow}
subtree is crashed as a group if it has too few nodes to determine the
height (even if the {\zhi} nodes would have added enough nodes).

\end{document}